\documentclass{aa}
\idline{1}{1}

\usepackage{graphicx}
\usepackage {amssymb}
\usepackage {mathrsfs}
\usepackage {txfonts}
\def\Teff  {T$_{\mbox {\scriptsize eff}}$}

\def\logg  {$\log g$}
\def\vt  {$v_t$}

\begin{document}

\title {Non-LTE abundances of Mg and K in
extremely metal-poor stars and the evolution of [O/Mg], [Na/Mg], 
[Al/Mg] and [K/Mg] in the Milky Way.
\thanks{Based on observations obtained with the ESO Very Large
Telescope at Paranal Observatory (Large Programme "First
Stars", ID 165.N-0276; P.I.: R. Cayrel.}
}

\author {
S.M. Andrievsky\inst{1,2}\and
M. Spite\inst{1}\and 
S.A. Korotin \inst{2}\and 
F. Spite\inst{1}\and
P. Bonifacio\inst{1,3,4}\and
R. Cayrel\inst{1}\and
P. Fran\c cois\inst{1}\and
V. Hill\inst{5}
}

\institute{
GEPI, Observatoire de Paris, CNRS, Universit\'e Paris Diderot; F-92125
Meudon Cedex, France, e-mail : {\tt monique.spite@obspm.fr}
\and
Department of Astronomy and Astronomical Observatory, Odessa National
University, Isaac Newton Institute of Chile, Odessa branch, Shevchenko
Park, 65014 Odessa, Ukraine, e-mail : {\tt scan@deneb1.odessa.ua}
\and
CIFIST Marie Curie Excellence Team
\and
Istituto Nazionale di Astrofisica, Osservatorio Astronomico di
Trieste, Via Tiepolo 11, I-34143 Trieste, Italy
\and 
Observatoire de la Cote dÕAzur, CNRS UMR6202, BP4229, 06304 Nice 
Cedex4, France
}

\date { }
\titlerunning{Relative abundances of light metals in EMP stars}
\authorrunning{Andrievsky}

\abstract
{}
{LTE abundances of light elements in extremely metal-poor
(EMP) stars have been previously derived from high quality spectra. 
New derivations, free from the NLTE effects,  will better constrain 
the models of the Galactic chemical
evolution and the yields of the very first supernovae.}
{The NLTE profiles of the magnesium and potassium lines have been
computed in a sample of 53 extremely metal-poor stars with a modified
version of the program MULTI and adjusted to the observed lines in order
to derive the abundances of these elements.}
{The NLTE corrections for magnesium and
potassium are in good agreement with the works found in the literature. 
 The abundances are slightly changed, reaching a better precision:  
 the scatter around the mean of the abundance ratios 
has decreased.
Magnesium may be used with confidence as reference element.
Together with previously determined NLTE abundances of sodium
and aluminum, the new ratios are displayed, for 
comparison, along the 
theoretical trends proposed by some models 
of the chemical evolution of
the Galaxy, using different models of supernovae.}
{}
\maketitle
%
\section{Introduction}

In the frame of the ESO Large Program "First stars, first
nucleosynthesis" Cayrel et al.  (\cite {CDS04}) and Bonifacio et al.
(\cite{BMS07}, \cite{BSC09}) have studied an homogeneous sample of
extremely metal-poor giants and turnoff stars.  For about fifty stars,
most of them with $\rm [Fe/H]<-3$, they determine the 
abundances of the elements from C to Zn, especially 
of the light metals Na, Mg, Al and K, in the early Galaxy.
These abundances are based on LTE computations of the equivalent
widths or of the line profiles.

After sodium and aluminum (Andrievsky et al., \cite{ASK07},
\cite{ASK08}) we present here non-LTE determinations of the magnesium
and potassium abundances by comparison of the observed and computed
line profiles and we discuss the evolution of the abundance
ratios [O/Mg], [Na/Mg], [Al/Mg] and [K/Mg] in the Galaxy.  Recently,
Takeda et al.  (\cite{TKM09}) have computed the non-LTE abundance of
potassium in the same sample of EMP stars, based on the equivalent
widths of the potassium lines given in Cayrel et al.  (\cite {CDS04}).
We compare our results with their determinations.

In Cayrel et al.  (\cite {CDS04}) and Bonifacio et al.  (\cite{BMS07},
\cite{BSC09}), iron had been first chosen as the main tracer of the
chemical evolution of the Galaxy. It had also been noted however,that
iron is perhaps not the best choice since there are several
processes that affect the yield of iron : Si burning in massive
SNe~II, explosive nucleosynthesis, mixing and fallback episodes and
possible late SN~Ia contributions.  Magnesium is, in principle, a better
choice since this element is mainly formed in massive SNe, its
production is dominated by hydrostatic carbon burning, and it is less
affected by explosive burning and fallback (e.  g. Woosley \& Weaver,
\cite{WW95}).  Among others, Shigeyama \& Tsujimoto (\cite{SS98})
recommended Mg rather than Fe as a reference element, following the
same logic.

  Cayrel et al.  (\cite{CDS04}) have recognised this too and have accordingly 
used magnesium as a reference element to compare
the observed ratios [X/Mg] to the ejecta of the massive supernovae
in the last part of the paper.
But it appeared that unexpectedly the relations [X/Mg] vs.  [Mg/H]
were more scattered than the corresponding relations [X/Fe] vs.
[Fe/H].  Also, in Bonifacio et al.  (\cite{BSC09}), the mean value 
of [Mg/Fe] differed significantly in giant
and turnoff stars.  As a consequence we suspected that both the
unexpected behaviour of the scatter of the ratios [X/Mg] 
and the different behaviour of dwarfs and giants
could be due at least partly to the neglect of the
departures from LTE.

\noindent - In Sect. 2 we present the characteristics of the atomic models used
for the non-LTE computations.\\  
\noindent - In Sect. 3 are presented the main
parameters of the analysis.\\  
\noindent - In Sect. 4 we discuss the ratios [Mg/Fe] and [K/Fe] in the
early Galaxy, and we compare the ratios [O/Mg], [Na/Mg], [Al/Mg] and
[K/Mg] at low metallicity to the predictions of the ejecta of
supernovae/hypernovae.  Finally for a complete halo-disk picture we
have also added the NLTE determinations of the abundances of Na, Mg, Al and
K in the disk from the "Gehren team" (Gehren et al.  \cite {GLS04},
\cite{GSZ06}, Mashonkina et al.  \cite{MZG08}, and Zhang et al.
\cite{ZGB06}), to study the evolution of these ratios in the Galaxy
and to compare this to the predictions of the models of the chemical
evolution of the Galaxy.

\begin {figure}[ht]
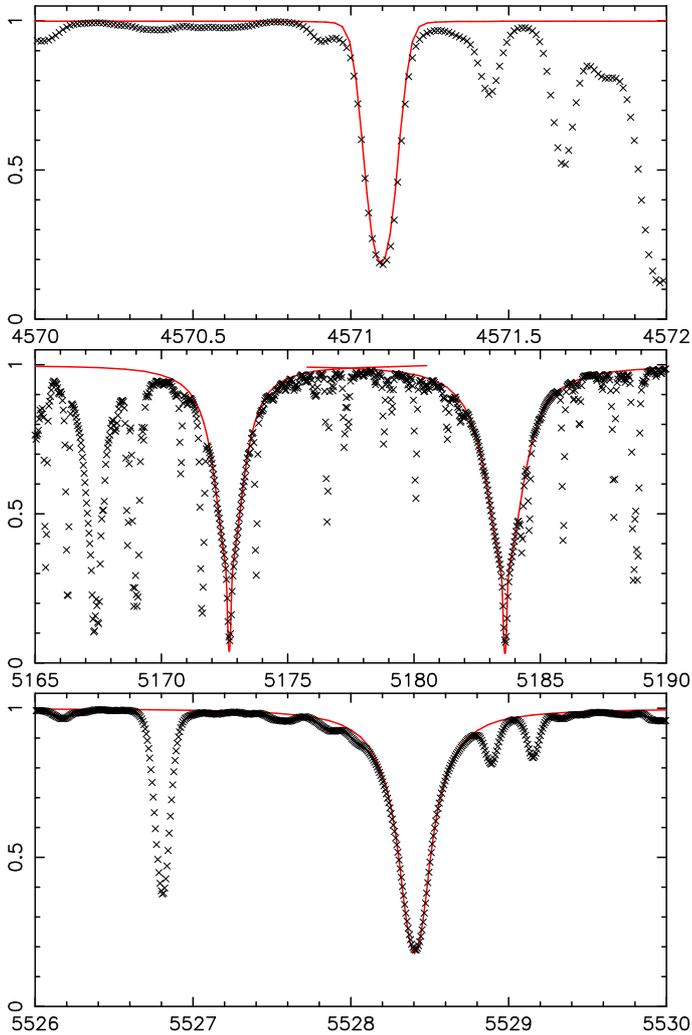

\begin {center}
\resizebox{\hsize}{!}
{\includegraphics[clip=true]{13223f1a.ps}}
\resizebox{\hsize}{!}
{\includegraphics[clip=true]{13223f1b.ps}}
\resizebox{\hsize}{!}
{\includegraphics[clip=true]{13223f1c.ps}}
\caption{ Profile fitting of the magnesium lines in the solar spectrum.}
\label{proSolMg}
\end {center}
\end {figure}

\begin {figure}[ht]
\begin {center}
\resizebox{\hsize}{!}
{\includegraphics[clip=true]{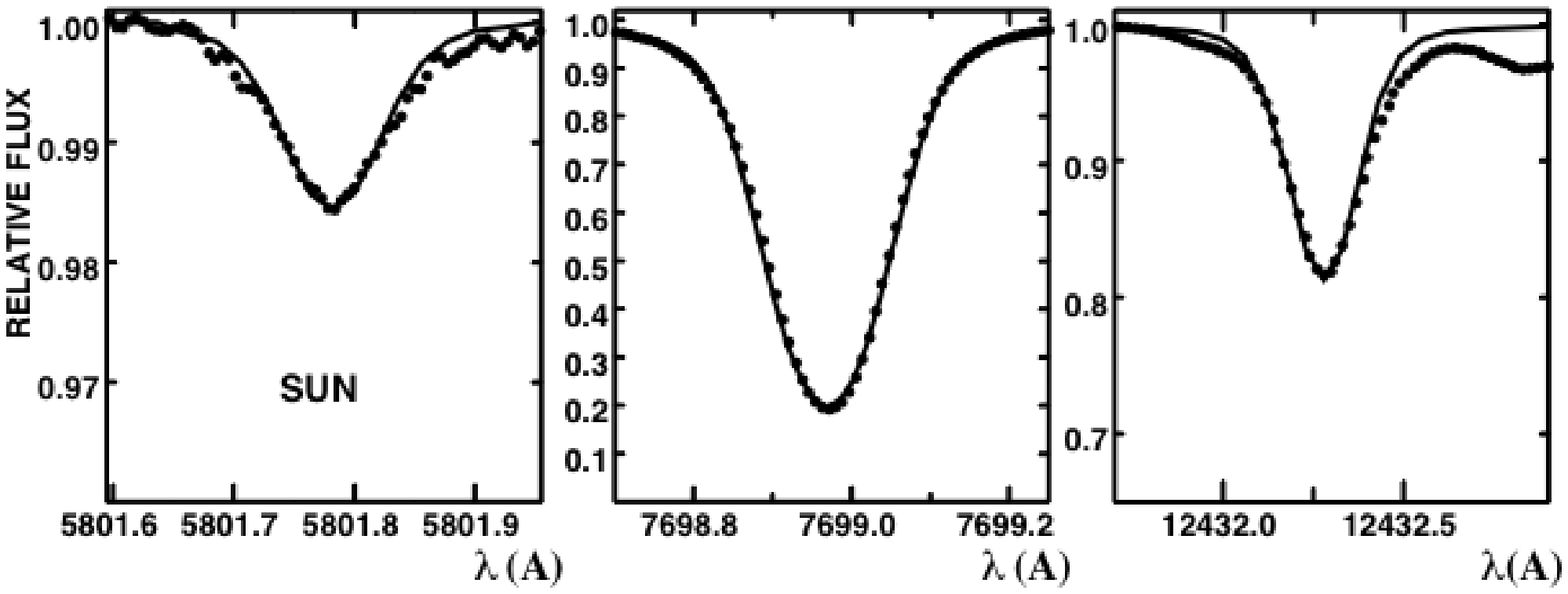}}
\caption{Profile fitting of the potassium lines in thesolar spectrum.}
\label {proSolK}
\end {center}
\end {figure}

\section{Atomic models and NLTE calculations}
\subsection{Magnesium}

The atomic model of magnesium used in this work is essentially the
same as the onedescribed in Mishenina et al.  (\cite{mskk04}).  This model
consists of 84 levels of \ion{Mg}{i}, 12 levels of \ion{Mg}{ii} and 
the ground level
of \ion{Mg}{iii} (Martin \& Zalubas, \cite{MZ80}, Biemont \& Brault
\cite{BB86}).  Within the described system the transitions between the
59 first levels of \ion{Mg}{i} and the ground level of \ion{Mg}{ii} have been
considered.  The detailed structure of the multiplets was ignored and
each LS multiplet was considered as a single term.  The fine structure
was taken into account only for the $\rm 3s3p^3P^0$ level (more 
details in Mishenina et al.  \cite{mskk04}).


\subsection{Potassium}

Our model of the \ion{K}{i} atom is based on the model
of Bruls et al.  (\cite{Bruls92}).  In our implementation we
considered in detail all the transitions between the first 20 levels
of the \ion{K}{i} and the ground level of the \ion{K}{ii}.  For the
level 4p2P we took into account the fine splitting.  The other levels
were considered as single levels.  Fifteen additional levels of
\ion{K}{i} and 7 levels of \ion{K}{ii} were used to allow the
conservation of the number of particles.  
The level energies are taken from Sugar \& Corliss (\cite{SC85}).  The
oscillator strengths of the bound-bound transitions are takenfrom Wiese et
al.  (\cite{WSM69}) as well as Biemont \& Grevesse (\cite{BG73}).  For the
resonance transitions we used data from Morton (\cite{Mor91}).  The
total number of bound-bound transitions considered in detail is 62.

Photoionization cross-sections from the ground level of potassium were
reported by Rahman-Attia et al.  (\cite{RJL86}).  For p-levels we took
ionization cross-sections published by Aymar et al.  (\cite{ALC76}).
For 2S, 2D, 2F levels the cross-sections calculated with the help of
quantum defect method (Hofsaess, \cite{Hof79}) were used.  For the
other levels, we used the hydrogen-like approximation.  The electron
collisional rates were estimated with the help of the Van Regemorter
(\cite{VReg62}) formula, while Allen's (\cite{All73}) formula was used
for forbidden transitions.  The collisional ionization from the ground
level was described using the corresponding formula from Sobelman et
al.  (\cite{SVY81}).  For the other levels we used Seaton's
(\cite{Sea62}) formula.  To take into account collisions of potassium
atoms with hydrogen atoms we applied the formula of Steenbock \& Holweger
(\cite{SH84}) with a correcting factor equal to 0.05.  A similar correcting
factor was used by Zhang et al.  (\cite{ZGB06}).

The principal difference between the LTE and NLTE-based abundances is
caused by an over-recombination on the first level of the \ion{K}{i}
atom in the atmospheres of the late-type stars.  This leads to an
increase of the equivalent widths of the \ion{K}{i} 766.5 and 769.9 nm
lines, and as a consequence the LTE abundances are overestimated.
Detailed discussion of this effect is given in Ivanova \& Shymansky
(\cite{IS00}).

\subsubsection{Fit of the solar spectrum}

To verify our adopted atomic models of magnesium and potassium we have
carried out the computation of the NLTE synthetic profiles of several
magnesium and potassium lines.  The Solar Flux Atlas of Kurucz
(\cite{Kur84}) in the visual and infra red range was used for this
purpose.

The NLTE profiles have been determined with the help of the modified
"MULTI" code of Carlsson (\cite{Carl86}).  The modifications are
described in Korotin et al.  (\cite{Kor99}), they include opacity 
sources from ATLAS9 (Kurucz \cite{Kur92}).
The Kurucz's (\cite{Kur96}) model of the solar atmosphere has been 
used in these computations.

The damping constants have been taken from the Vienna Atomic Line Data 
Base (VALD, Piskunov et al., \cite{PKR95}).  
The correction $\rm \Delta log C_{6}$ found by Mishenina et al.
(\cite {mskk04}) has been applied for the computation of the magnesium
lines.  The NLTE profiles of the Mg I lines computed with $\rm log
(Mg/H)=7.57$ agree well with the solar spectrum (see Fig.
\ref{proSolMg} in good agreement with the determinations of Grevesse
\& Sauval (\cite{GS00}), and of Shimanskaya (\cite{SMS00}) who found
$\rm log (Mg/H)_{\odot}=7.58$.  The potassium lines are also well
reproduced by our calculations (see Fig.  \ref{proSolK}) with a
potassium abundance $\rm log (K/H)_{\odot}=5.11$, in good agreement
with the value adopted by Zhang et al.  (\cite{ZGB06}), and very close to
the meteoritic potassium abundance (Lodders, \cite{Lod03}).

\section{Analysis of the star sample}

\begin {table*}[th]
\caption {Adopted model and potassium abundance for our sample of
stars.  In the last column the letter m indicates that the giant has
been found ``mixed'' (see text).}
\label {tabstars}
\begin {center}
\begin{tabular}{rlc@{~~~~~}l@{ }c@{ }c@{~~~~~~~}c@{ }c@{ }c@{ }c@{~~~~~}c@{ }c@{ }c@{ }c@{ }c@{ }c@{ }c@{ }c}
  & & $\rm T_{eff}$& log &$\rm v_{t}$ & & LTE & NLTE & NLTE & NLTE &
  LTE & NLTE & NLTE & NLTE & \\
~ & star & (K) &$~~g~~$&$\rm (km s^{-1})$ & [Fe/H] &$\rm\epsilon(Mg)$&
$\rm\epsilon(Mg)$&[Mg/H]&[Mg/Fe]&$\rm\epsilon(K)$&
$\rm\epsilon(K)$&[K/H]&[K/Fe]& Rem\\
\hline
&\bf{ GIANTS}\\
01 & HD 2796       & 4950&  1.5&  2.1&  -2.47&  5.54&  5.74&  -1.84&  0.63&       3.25&  2.90&  -2.22&  0.25&   m\\
02 & HD 122563     & 4600&  1.1&  2.0&  -2.82&  5.29&  5.39&  -2.19&  0.63&       2.78&  2.57&  -2.55&  0.27&   m\\
03 & HD 186478     & 4700&  1.3&  2.0&  -2.59&  5.55&  5.72&  -1.86&  0.73&       3.10&  2.85&  -2.27&  0.32&   m\\
04 & BD +17:3248   & 5250&  1.4&  1.5&  -2.07&  6.00&  6.19&  -1.39&  0.68&       3.77&  3.35&  -1.77&  0.30&   m\\
05 & BD -18:5550   & 4750&  1.4&  1.8&  -3.06&  4.99&  5.14&  -2.44&  0.62&       2.63&  2.45&  -2.67&  0.39&    \\
06 & CD -38:245    & 4800&  1.5&  2.2&  -4.19&  3.66&  4.02&  -3.56&  0.63&         - &    -&     -  &   - &    m\\
07 & BS 16467-062  & 5200&  2.5&  1.6&  -3.77&  3.95&  4.24&  -3.34&  0.43&       1.85&  1.70&  -3.42&  0.35&    \\
08 & BS 16477-003  & 4900&  1.7&  1.8&  -3.36&  4.76&  4.86&  -2.72&  0.64&       2.30&  2.04&  -3.08&  0.28&    \\
09 & BS 17569-49   & 4700&  1.2&  1.9&  -2.88&  5.36&  5.47&  -2.11&  0.77&       2.84&  2.59&  -2.53&  0.35&   m\\
10 & CS 22169-035  & 4700&  1.2&  2.2&  -3.04&  4.76&  4.92&  -2.66&  0.38&       2.55&  2.37&  -2.75&  0.29&   m\\
11 & CS 22172-002  & 4800&  1.3&  2.2&  -3.86&  3.89&  4.19&  -3.39&  0.47&       1.85&  1.67&  -3.45&  0.41&    \\
12 & CS 22186-025  & 4900&  1.5&  2.0&  -3.00&  5.05&  5.19&  -2.39&  0.61&       2.68&  2.47&  -2.65&  0.35&   m\\
13 & CS 22189-009  & 4900&  1.7&  1.9&  -3.49&  4.29&  4.47&  -3.11&  0.38&       2.20&  2.06&  -3.06&  0.43&    \\
14 & CS 22873-055  & 4550&  0.7&  2.2&  -2.99&  5.11&  5.24&  -2.34&  0.65&       2.60&  2.39&  -2.73&  0.26&   m\\
15 & CS 22873-166  & 4550&  0.9&  2.1&  -2.97&  5.27&  5.44&  -2.14&  0.83&       2.71&  2.52&  -2.60&  0.37&   m\\
16 & CS 22878-101  & 4800&  1.3&  2.0&  -3.25&  4.85&  5.02&  -2.56&  0.69&       2.35&  2.16&  -2.96&  0.29&   m\\
17 & CS 22885-096  & 5050&  2.6&  1.8&  -3.78&  4.13&  4.39&  -3.19&  0.59&       1.80&  1.72&  -3.40&  0.38&    \\
18 & CS 22891-209  & 4700&  1.0&  2.1&  -3.29&  4.76&  4.89&  -2.69&  0.60&       2.35&  2.15&  -2.97&  0.32&   m\\
19 & CS 22892-052* & 4850&  1.6&  1.9&  -3.03&  4.90&  5.06&  -2.52&  0.51&       2.55&  2.34&  -2.78&  0.25&    \\
20 & CS 22896-154  & 5250&  2.7&  1.2&  -2.69&  5.22&  5.52&  -2.06&  0.63&       2.88&  2.62&  -2.50&  0.19&    \\
21 & CS 22897-008  & 4900&  1.7&  2.0&  -3.41&  4.54&  4.75&  -2.83&  0.58&       2.20&  2.02&  -3.10&  0.31&    \\
22 & CS 22948-066  & 5100&  1.8&  2.0&  -3.14&  4.81&  4.99&  -2.59&  0.55&       2.55&  2.37&  -2.75&  0.39&   m\\
23 & CS 22949-037* & 4900&  1.5&  1.8&  -3.97&  5.07&  5.16&  -2.42&  1.55&       1.50&  1.32&  -3.80&  0.17&   m\\
24 & CS 22952-015  & 4800&  1.3&  2.1&  -3.43&  4.20&  4.51&  -3.07&  0.36&       2.15&  1.99&  -3.13&  0.30&   m\\
25 & CS 22953-003  & 5100&  2.3&  1.7&  -2.84&  5.03&  5.24&  -2.34&  0.50&       2.60&  2.42&  -2.70&  0.14&    \\
26 & CS 22956-050  & 4900&  1.7&  1.8&  -3.33&  4.59&  4.99&  -2.59&  0.74&       2.20&  2.09&  -3.03&  0.30&    \\
27 & CS 22966-057  & 5300&  2.2&  1.4&  -2.62&  5.39&  5.69&  -1.89&  0.73&       2.95&  2.77&  -2.35&  0.27&    \\
28 & CS 22968-014  & 4850&  1.7&  1.9&  -3.56&  4.27&  4.54&  -3.04&  0.52&       1.83&  1.77&  -3.35&  0.21&    \\
29 & CS 29491-053  & 4700&  1.3&  2.0&  -3.04&  5.09&  5.24&  -2.34&  0.70&       2.69&  2.42&  -2.70&  0.34&   m\\
30 & CS 29495-041  & 4800&  1.5&  1.8&  -2.82&  5.23&  5.49&  -2.09&  0.73&       2.91&  2.63&  -2.49&  0.33&    \\
31 & CS 29502-042  & 5100&  2.5&  1.5&  -3.19&  4.74&  5.06&  -2.52&  0.67&       2.26&  2.07&  -3.05&  0.14&    \\
32 & CS 29516-024  & 4650&  1.2&  1.7&  -3.06&  5.25&  5.34&  -2.24&  0.82&         - &    -&     -  &   - &     \\
33 & CS 29518-051  & 5200&  2.6&  1.4&  -2.69&  5.27&  5.56&  -1.02&  0.67&       2.94&  2.67&  -2.45&  0.24&   m\\
34 & CS 30325-094  & 4950&  2.0&  1.5&  -3.30&  4.81&  5.04&  -2.54&  0.76&       2.65&  2.42&  -2.70&  0.60&    \\
35 & CS 31082-001  & 4825&  1.5&  1.8&  -2.91&  5.21&  5.49&  -2.09&  0.82&       2.80&  2.47&  -2.65&  0.26&    \\
\hline 
&\bf{ TURNOFF}\\ 
1  &BS~16023--046  & 6360 & 4.5 & 1.4&  -2.97&  4.67&  4.97&  -2.61&  0.36\\ 
2  &BS~16968--061  & 6040 & 3.8 & 1.5&  -3.05&  4.82&  5.12&  -2.46&  0.59\\ 
3  &BS~17570--063  & 6240 & 4.8 & 0.5&  -2.92&  4.74&  5.06&  -2.52&  0.40\\ 
4  &CS~22177--009  & 6260 & 4.5 & 1.2&  -3.10&  4.70&  5.11&  -2.47&  0.63\\ 
5  &CS~22888--031  & 6150 & 5.0 & 0.5&  -3.28&  4.51&  5.04&  -2.54&  0.74\\ 
6  &CS~22948--093  & 6360 & 4.3 & 1.2&  -3.43&  4.33&  4.72&  -2.86&  0.57\\ 
7  &CS~22953--037  & 6360 & 4.3 & 1.4&  -2.89&  5.05&  5.34&  -2.24&  0.65\\ 
8  &CS~22965--054  & 6090 & 3.8 & 1.4&  -3.04&  4.79&  5.16&  -2.42&  0.62\\ 
9  &CS~22966--011  & 6200 & 4.8 & 1.1&  -3.07&  4.72&  5.01&  -2.57&  0.50\\
10 &CS~29499--060  & 6320 & 4.0 & 1.5&  -2.70&  5.07&  5.44&  -2.14&  0.56\\
11 &CS~29506--007  & 6270 & 4.0 & 1.7&  -2.91&  4.95&  5.33&  -2.25&  0.66\\
12 &CS~29506--090  & 6300 & 4.3 & 1.4&  -2.83&  5.02&  5.36&  -2.22&  0.61\\
13 &CS~29518--020  & 6240 & 4.5 & 1.7&  -2.77&  4.87&  5.31&  -2.27&  0.50\\
14 &CS~29518--043  & 6430 & 4.3 & 1.3&  -3.24&  4.57&  4.94&  -2.64&  0.60\\
15 &CS~29527--015  & 6240 & 4.0 & 1.6&  -3.55&  4.46&  4.86&  -2.72&  0.83\\
16 &CS~30301--024  & 6330 & 4.0 & 1.6&  -2.75&  5.11&  5.33&  -2.25&  0.50\\
17 &CS~30339--069  & 6240 & 4.0 & 1.3&  -3.08&  4.68&  5.04&  -2.54&  0.54\\
18 &CS~31061--032  & 6410 & 4.3 & 1.4&  -2.58&  5.22&  5.44&  -2.14&  0.44\\
\hline   
\end {tabular}  
\end {center}
* An asterisk after the name of the star means that the star is carbon-rich.   
\end {table*}

\subsection{Sample of stars. Observations and reduction} 

The sample of stars and the observational data are the same as
discussed in Cayrel et al.  (\cite{CDS04}).  The observations were
performed with the high resolution spectrograph UVES at the VLT
(Dekker et al., \cite{DDK00}).  The resolving power in the region of
the magnesium lines is $R \approx 45,000$ and it is $R \approx 
41,000$ in the region of the potassium lines.  The S/N ratio is generally $
\approx 120/pix$ with about 5 pixels per resolution element. In 
this region of the potassium lines, residual fringes limit the
precision of the measurements and in most of the stars a line with
an equivalent width less than 3 m\AA~ cannot be detected.

The spectra have been reduced using the UVES context (Ballester et al.
\cite{BMB00}).

\subsection{Atmospheric parameters}
The parameters of the atmosphere of the stars \Teff, \logg, \vt ~and
[Fe/H] are given in Table \ref{tabstars}, and are quoted from the
LTE analyses of Cayrel et al.  (\cite{CDS04}) and Bonifacio et al.
(\cite{BMS07}, \cite{BSC09}).  An "m" in the last column of the table
means that the atmosphere of the star (giant) has been found "mixed"
with the deep hydrogen burning layer by Spite et al.  (\cite{SCP05}, 
\cite{SCH06}).
 For the determination of the relative LTE and NLTE
abundances, we adopted in this table the solar values 
$\rm log(Mg/H)_{\odot}=7.58$ and $\rm log (K/H)_{\odot}=5.12$ following
Grevesse \& Sauval (\cite{GS00}) for a better homogeneity with the
previous LTE determinations (Cayrel et al. \cite{CDS04}).

\subsection{LTE abundances of K and Mg}
As a comparison, we give  in Table \ref{tabstars} new abundances
of magnesium and potassium computed with the LTE hypothesis.  These
new determinations have been obtained by fitting the  
 synthetic spectra with the observed profiles.  

For the abundance of potassium the result is
sometimes slightly different from the value published in Cayrel et
al.  (\cite{CDS04}) because, for example, a different position of the 
continuum has been adopted.

However, the difference is larger for magnesium.  In Cayrel et
al.  (\cite{CDS04}) 
 the equivalent widths of magnesium lines in giants have been often 
 underestimated: the lines are often 
strong and the wings had been neglected.  This error has been corrected
in Bonifacio et al.  (\cite{BSC09}).  The correction is negligible for
the most magnesium-poor giants, but in in the other (less Mg-poor) 
giants, the difference is about 0.15 dex.

\subsection{NLTE abundances}
To compute NLTE profiles of the magnesium and potassium lines, we have
used Kurucz's models (ATLAS9) without overshooting (Kurucz
\cite{Kur93}).  We have checked on some typical stars that the use of
MARCS models (Gustafsson et al., \cite{GEE08}) as in Cayrel et al.
(\cite{CDS04}) would not make a significant difference.

The NLTE corrections depend on the effective temperature and gravity
of the model, as well as on the element abundance itself.  The latter
circumstance strongly suggests that the abundances must be derived
individually for each star using complete NLTE computation.  The use
of published NLTE corrections as a function \Teff ~and \logg ~can
introduce some errors in the derived abundances if the abundances are
different.   For example in metal-poor stars in the range $\rm-2.0
<[Fe/H]<-3.0$ the NLTE corrections for the potassium abundance are
within the range 0.2-0.5 dex.  But in the extremely metal-poor stars
($\rm [Fe/H]<-3.0$) the potassium line is very weak (the equivalent
width of the 7699~\AA~  line is always less than 25 m\AA~ and often less
than 10 m\AA); the line is thus formed rather deeply in the atmosphere
where collisions are important and finally the NLTE correction at
these metallicities is small. As a consequence, the use of a uniform NLTE
correction (computed by Ivanova \& Shymansky for a metallicity of 
-2.0),
has led Cayrel et al.  (\cite{CDS04}) to underestimate the
potassium abundance at very low metallicity.

\begin {figure}[h]
\begin {center}
\resizebox{\hsize}{!}
{\includegraphics[clip=true]{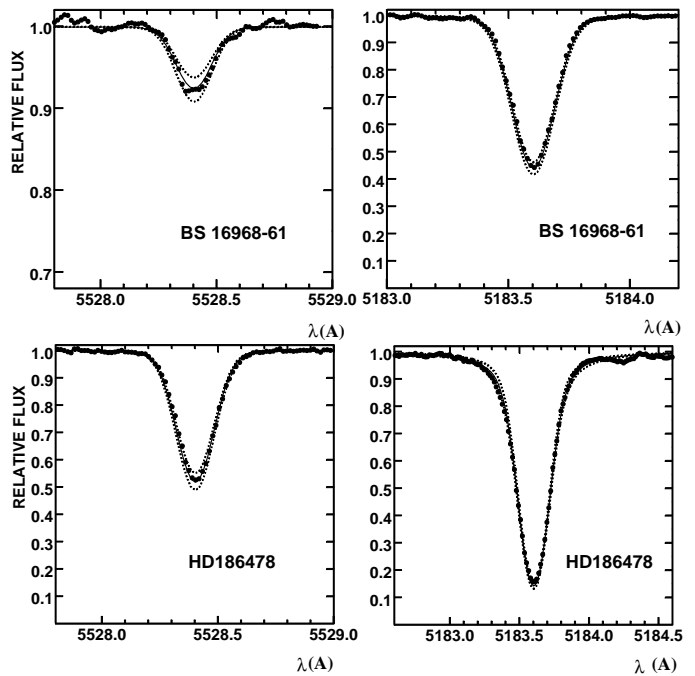}}
\caption{Magnesium. Profile fitting for two metal-poor stars with different 
effective  temperature, with similar metallicity :$\rm[Fe/H] \approx -2.8$.  
One is a turnoff star (BS~16968-61), the other a giant (HD~186478). The
Mg abundance was varied by 0.10 dex (dotted lines).}
\label {prStarMg}
\end {center}
\end {figure}

\begin {figure}[ht]
\begin {center}
\resizebox{\hsize}{!}
{\includegraphics[clip=true]{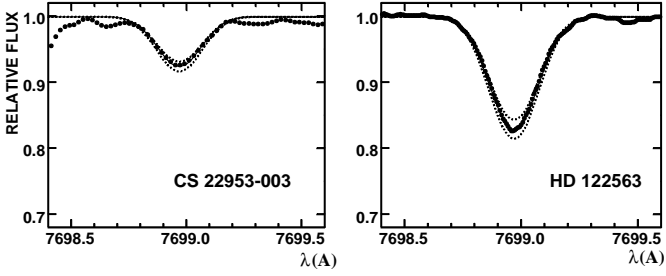}}
\caption{Potassium. Profile fitting for two metal-poor giants, both with $\rm
[Fe/H] \approx -2.8$ but with different effective temperature.  The
best-fit K abundance was varied by 0.05 dex.}
\label {prStarK}
\end {center}
\end {figure}

The profile fitting for two metal-poor stars is displayed in Fig.
\ref{prStarMg} and \ref{prStarK}.
The NLTE magnesium and potassium abundances in our
program stars are listed in Table \ref{tabstars}. In this table, we
give also as a comparison the abundance of these elements computed
independently with the LTE approximation (and without any correction).
The potassium abundance can be computed only in giant stars, the 
lines are not visible in the turnoff stars.

\section{Discussion} \label{discuss}

\subsection{Abundance of magnesium and potassium in the early Galaxy}

As recalled in Cayrel et al.  (\cite {CDS04}), Mg is formed during
hydrostatic carbon burning and explosive neon burning, K during
explosive oxygen burning; their
abundances are therefore related to the importance of both the
hydrostatic and explosive phases.

\begin {figure}[ht]
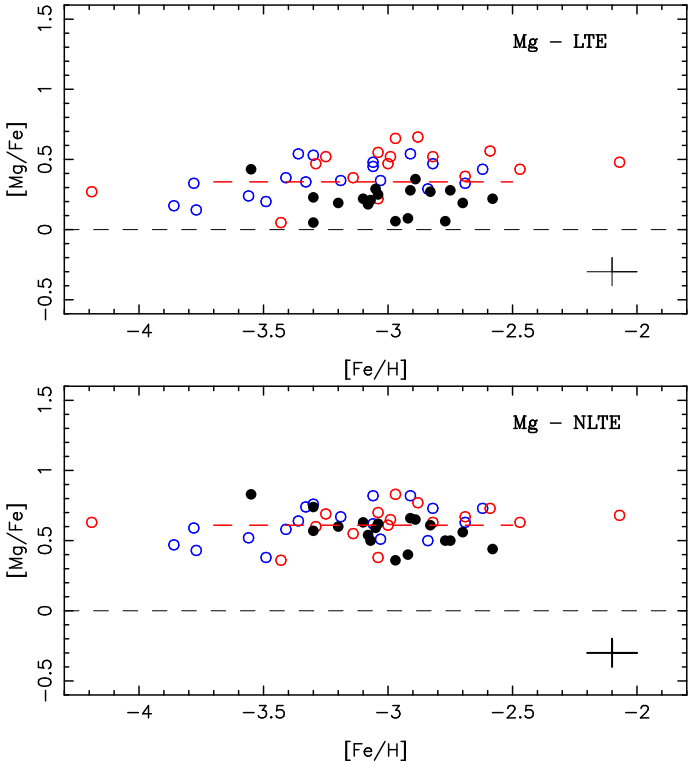

\begin {center}
\resizebox{\hsize}{!}
{\includegraphics[clip=true]{13223f5a.ps}}
\resizebox{\hsize}{!}
{\includegraphics[clip=true]{13223f5b.ps}}
\caption{ [Mg/Fe] vs. [Fe/H] in the early Galaxy.
Open circles (red and blue) stand for "mixed" and "unmixed" giants (see Spite 
et al. \cite{SCP05}, \cite{SCH06}, \cite{SCS06}), black filled symbols 
for turnoff stars.}
\label {abfemg}
\end {center}
\end {figure}

In Fig.  \ref{abfemg} we compare the LTE abundances of magnesium in
our sample of EMP dwarfs and giants (Bonifacio et al., \cite{BSC09})
with the new NLTE determinations.  The carbon-rich and peculiar star 
CS 22949-037 has been discarded.

\begin {figure}[ht]
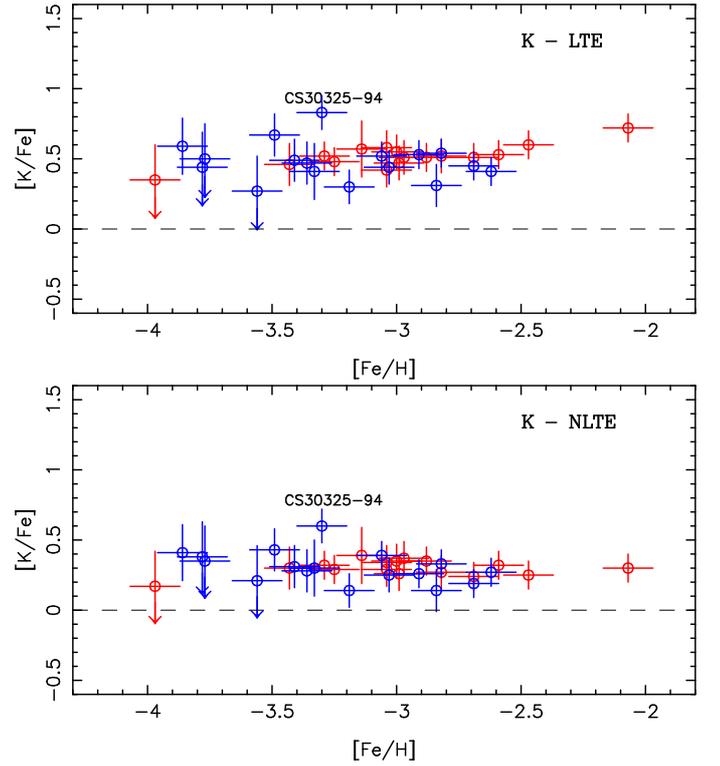

\begin {center}
\resizebox{\hsize}{!}
{\includegraphics[clip=true]{13223f6a.ps}}
\resizebox{\hsize}{!}
{\includegraphics[clip=true]{13223f6b.ps}}
\caption{ [K/Fe] vs. [Fe/H] in the early Galaxy.
Open circles (red and blue) stand for "mixed" and "unmixed" giants 
respectively (see Spite 
et al. \cite{SCP05}, \cite{SCH06}). The potassium abundance cannot 
be measured in the turnoff stars, the lines are too weak.}
\label {abfek}
\end {center}
\end {figure}

The trend of the [Mg/Fe] ratio vs.  metallicity at low metallicity is
not changed, it remains a plateau.  But the scatter of [Mg/Fe] is
smaller when NLTE effects are taken into account: 0.13 from NLTE
computations and 0.16 from LTE computation.  The difference between
dwarfs and giants is strongly reduced and is not significant any more.  The
mean value of [Mg/Fe] in the interval $\rm -4<[Fe/H]<-2.5$ is now
+0.61dex (it was +0.34 from LTE computations). This value of [Mg/Fe]
is now similar to the mean value of [O/Fe] previously deduced (Cayrel 
et al. 2004) from 
the forbidden oxygen line (not sensitive to NLTE effects).

For the determination of the mean value of the ratios [Mg/Fe], their trend 
and their scatter around the trend, the carbon-rich star  
CS~22949-037 (showing also strong abundance anomalies of light 
elements) has been discarded. The star has been as well excluded from
all computations and/or figures related to Mg abundances.


When NLTE is taken into account, the values of the abundance ratios
[K/Fe] are slightly
decreased. The scatter of [K/Fe] is also slightly smaller.  The mean
value of [K/Fe] in the interval $\rm -4<[Fe/H]<-2.5$ is about
+0.33 dex.  This value is very close to the value found by Takeda et
al.  (\cite{TKM09}).  We have checked that the small differences
between the two determinations reflect the fact that we have
determined the potassium abundance by a direct fit of the profiles
whereas Takeda et al.  have used the equivalent widths given in Cayrel
et al.  (\cite {CDS04}).

We have estimated star by star the error on [K/Fe] in our sample of
stars.  This error is dominated by the measurement error (equivalent
width or profile) and it increases significantly when the metallicity
decreases: at low metallicity the potassium lines becomes very weak
(less than 10 m\AA ~and often less than 5 m\AA) in a region where the
S/N ratio is only about 100 and where the position of the continuum
is not very precise because of residual fringes.  In consequence, 
as  shown in our  Fig.  \ref{abfek} we
do not confirm the suggestion of Takeda et al. (\cite{TKM09}) of
"a marginal sign of decline toward a further lower [Fe/H]".

It is interesting to note that the K-rich star CS~30325-94 (LTE, see
Cayrel et al.  \cite{CDS04}) remains K-rich (and with a small error)
in the NLTE analysis (see Fig.  \ref{abfek}) and it is also Sc-rich.
In the stellar sample of Zhang \& Zhao (\cite{ZZ05}), the most
potassium-rich star HD 195636 (with [Fe/H]=--3.3, and [K/Fe]=+1.35) is
also very scandium-rich.  The production of these elements is not
unanimously ascribed to definite sites; there could be a link between
the formation processes of these elements in massive stars.
However, the scandium-rich stars are not always potassium-rich (see 
for example CS 22885-096 in our sample).

\subsection {Abundances of O and light metals relative to Mg in the 
early Galaxy}

\begin {figure}[ht]
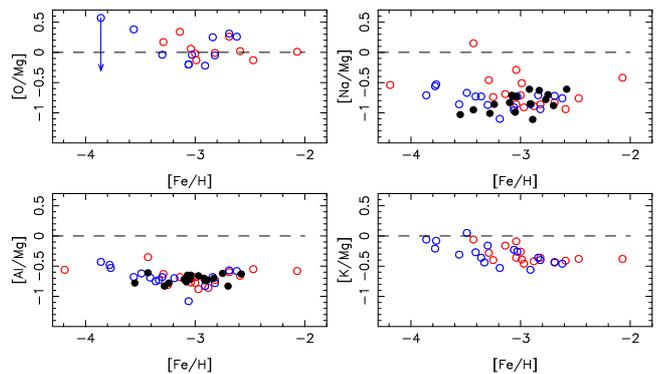

\begin {center}
\resizebox  {4.2cm}{2.4cm}
{\includegraphics {13223f7a.ps} }
\resizebox  {4.2cm}{2.4cm}
{\includegraphics {13223f7b.ps} }
\resizebox  {4.2cm}{2.4cm}
{\includegraphics {13223f7c.ps} }
\resizebox  {4.2cm}{2.4cm}
{\includegraphics {13223f7d.ps} }
\caption{Abundance of O, Na, Al, and K relative to Mg in the early
Galaxy.  The abundances of all these elements have been computed
taking into account the NLTE effect.  Since the measured oxygen lines are
forbidden lines they are free from NLTE effects.}
\label {el-mg}
\end {center}
\end {figure}

Oxygen:

The magnesium abundances listed in Table \ref{tabstars} can be used 
as a
normalization factor for other elements.  It is interesting to compute the
ratio [O/Mg] using this new NLTE value of the magnesium abundance :
since the abundance of oxygen has been determined from the forbidden
oxygen line, is free from NLTE effects.  The corresponding
plots are shown in Fig.  \ref{el-mg}. 

Sodium:

As found in Andrievsky et al.
(\cite {ASK07}), the "mixed" stars are often enriched in sodium.  This
reflects an internal mixing inside these giant stars.  Therefore the
mixed stars should not be used to determine the ratio [Na/Mg] in the
early Galaxy. Note that the star CS~22952-15 (a mixed star with 
[Fe/H]=--3.43) is very sodium-rich, but its ratio [Al/Fe] is 
normal (see Andrievsky et al., \cite{ASK08}). However its ratio 
[Al/Mg] is also rather high. The explanation is that 
this mixed star is Na-rich and Mg-poor.

Mean value and scatter:

Since the  abundances of O, Na, Al and K relative to Mg are rather 
flat in their central part, we can define a mean value in this
central interval, say  $\rm -3.6<[Fe/H]< -2.5$. The mean values of 
these flat parts are : 

 ~$\rm [O/Mg] \approx 0.1$, $\rm [Na/Mg] \approx
-0.8$, $\rm [Al/Mg] \approx -0.7$ and $\rm [K/Mg] \approx -0.3$.  

For  
the same sample of stars in the same interval of metallicity 
($\rm -3.6<[Fe/H]<-2.5$),  
the scatter around the mean  can also be computed, the values are :

$\rm \sigma_{[O/Fe]}=0.18$ but $\rm \sigma_{[O/Mg]}=0.20$, 
$\rm \sigma_{[Na/Fe]}=0.12$ but $\rm \sigma_{[Na/Mg]}=0.14$, 
$\rm \sigma_{[Al/Fe]}= 0.11, \sigma_{[Al/Mg]}=0.11$,  
$\rm \sigma_{[K/Fe]}=0.09$ but $\rm \sigma_{[K/Mg]}=0.15$.

The large  scatter found for O, is due to the 
reduced number of measurements. 

The values of the scatter obtained for the NLTE abundance ratios are 
lower than for the LTE abundances. However, it appears that, 
as already observed in LTE (Cayrel et al., 2004),
the scatter of the abundance ratios  is larger when
Mg is chosen as the reference element instead of Fe.  Al is an
exception.  

This fact is surprising: it would have been expected that the 
abundances of Na and K produced by (mainly hydrostatic) nuclear 
processes (similar to those producing Mg) 
would bear more resemblance to the Mg abundance than to the Fe 
abundance (produced quite differently); the scatter of [Na/Mg] and
[K/Mg] would then be smaller than the scatter of [Na/Fe] and [K/Fe],
but the observations show the contrary.

\subsection{Comparison to the predictions of the ejecta of supernovae}

\begin {figure}[ht]
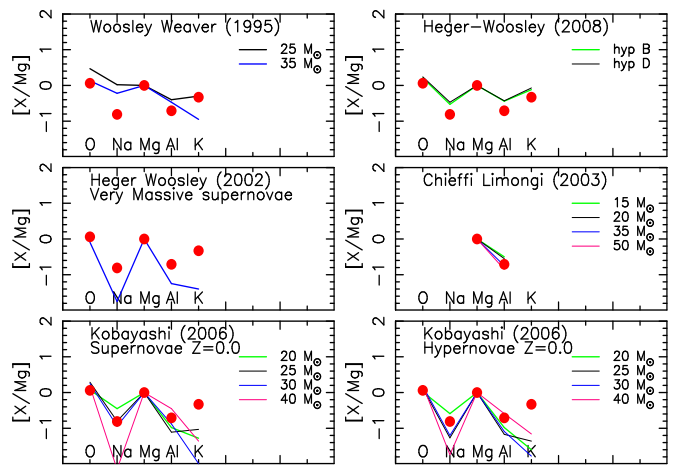

\begin {center}
\resizebox {4.3cm}{2.0cm}
{\includegraphics {13223f8a.ps} }
\resizebox  {4.3cm}{2.0cm}
{\includegraphics {13223f8b.ps} }
\resizebox  {4.3cm}{2.0cm}
{\includegraphics {13223f8c.ps} }
\resizebox  {4.3cm}{2.0cm}
{\includegraphics {13223f8d.ps} }
\resizebox  {4.3cm}{2.0cm}
{\includegraphics {13223f8e.ps} }
\resizebox  {4.3cm}{2.0cm}
{\includegraphics {13223f8f.ps} }
\caption{ Comparison of the new abundance ratios to the predictions 
of supernovae or hypernovae. }
\label {compSN}
\end {center}
\end {figure}

It is interesting to compare the abundance ratios [O/Mg], [Na/Mg],
[Al/Mg] and [K/Mg] to the predictions of the ejecta of metal-poor
supernovae or hypernovae.  
In Fig.  \ref{compSN}  the observed ratios are compared to the
predictions of Woosley \& Weaver (\cite {WW95}), of Heger \& Woosley (\cite
{HW08}, \cite{HW02}), of Chieffi \& Limongi (\cite{CL03}), and Kobayashi et al.
for supernovae and hypernovae (Kobayashi et al., \cite{KUN06}).

The quantity of potassium ejected is generally underestimated by the
models of supernovae.  The best agreement is obtained by
the predictions of Heger \& Woosley (\cite {HW08}) 
with the hypotheses B and D (see the paper), although the abundance
of sodium and aluminum are in this case a little overestimated. The 
ratio Al/Mg is also well represented in the model of Chieffi \& Limongi 
(\cite{CL03}). Unfortunately there are no predictions for the 
abundance of O, Na and K in this model. The models of Kobayashi et 
al. (\cite {KUN06}) 
(supernovae and hypernovae with Z=0.0) seem also able to represent 
the relative abundances of O, Na, Mg and Al, but here again the 
production of potassium is strongly underestimated.

\subsection{Variation of [O/Mg], [Na/Mg], [Al/Mg] and [K/Mg] in the Galaxy}

\begin {figure}[ht]
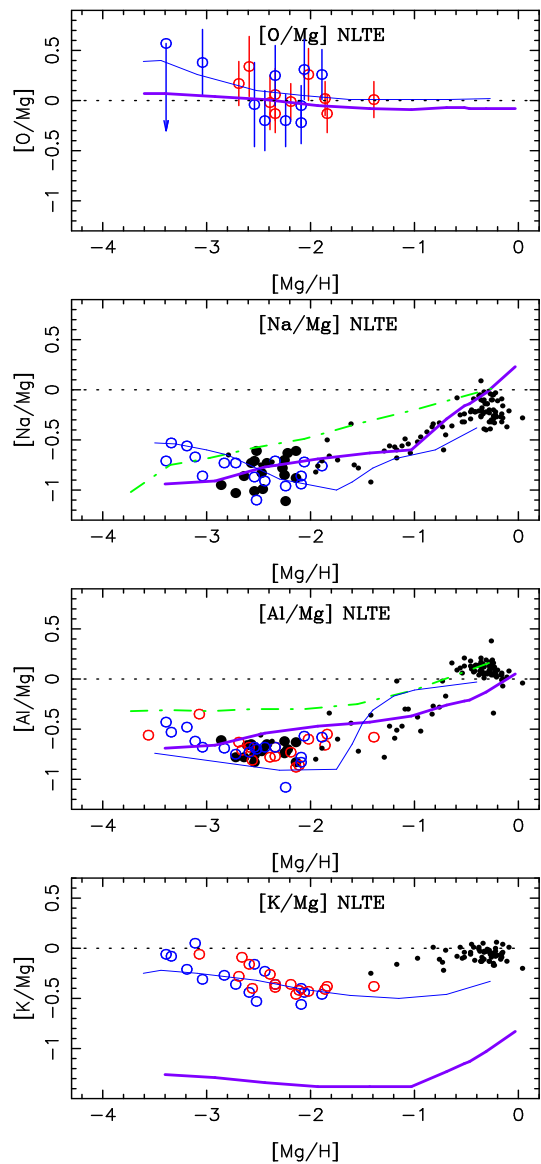

\begin {center}
\resizebox {7.cm}{3.8cm}
{\includegraphics {13223f9a.ps} }
\resizebox {7.cm}{3.8cm}
{\includegraphics {13223f9b.ps} }
\resizebox {7.cm}{3.8cm}
{\includegraphics {13223f9c.ps} }
\resizebox {7.cm}{3.8cm}
{\includegraphics {13223f9d.ps} }
\caption{Variation of [O/Mg], [Na/Mg], [Al/Mg], [K/Mg] vs.  [Mg/H] in
the Galaxy.  The red and blue open circles represent our measurements
for mixed and unmixed giants (see the electronic version of A\&A for a
color version of this figure), the large black dots represent the
turnoff stars of our sample and the small black dots are from Gehren
et al.  (\cite {GLS04}, \cite{GSZ06}), Mashonkina et al.
(\cite{MZG08}), and Zhang et al.  (\cite{ZGB06}), for the disk and
halo stars.  The thin solid line is the prediction of Fran\c cois et al.
(\cite {FMC04}) for O and K and of Cescutti et al. (private 
communication) for Na and Mg. The dashed dotted lines are the
predicitions of Goswami \& Prantzos (\cite {GP00}) for Na and Al.
These predictions are based (with  occasional adjustments) 
on the models of supernovae ejecta of
Woosley \& Weaver (\cite{WW95}).  The thick solid lines are the
predictions of Kobayashi et al.  (\cite{KUN06}) based (without 
adjustments) on new models of supernovae and hypernovae.  
} 

\label {abmg-naalk} 
\end {center} 
\end {figure}

In Fig. \ref{abmg-naalk} we present the variation of [O/Mg], [Na/Mg], [Al/Mg] 
and [K/Mg] in the Galaxy as a function of [Mg/H] where Na, Al, Mg and K 
have been computed taking into account the NLTE effects (the oxygen 
abundance deduced from forbidden lines is not affected by departure 
from LTE).

As already noted, [Mg/H] should be a better index of time, since magnesium 
is only
formed in massive SN~II with a short lifetime, unlike iron formed in
SN~II or SN~I of various masses.

We have added in the figure the determinations of Gehren et al. (\cite 
{GLS04}, \cite{GSZ06}), of Mashonkina et al. (\cite{MZG08}), and 
Zhang et al. (\cite{ZGB06}) in the Galactic disk and halo. All 
these determinations have been done including the NLTE effects 
on the line profiles.

As seen in Fig. 9, the abundance ratios [Na/Mg], [Al/Mg] and [K/Mg] in the
Galaxy decrease with [Mg/H] from [Mg/H]= 0 to [Mg/H]=--2, then
are almost constant between [Mg/H]=--2 and [Mg/H]=--2.8, and finally, at
lower metallicity, the ratios [Na/Mg], [Al/Mg], and [K/Mg] and maybe
also [O/Mg] seem to increase when [Mg/H] decreases.  (The error on
[O/Mg] in the two most magnesium-poor stars is large and thus this
tendency is not firmly established for oxygen.)

In Fig.  \ref{abmg-naalk}  the evolutions of the
abundance ratios [O/Mg], [Na/Mg], [Al/Mg] and [K/Mg] in the Galaxy 
are also drawn,
following Kobayashi et al.  (\cite{KUN06}), Fran\c cois et al.  (\cite
{FMC04}) and for Na and Al, Goswami \& Prantzos (\cite {GP00}).  

Kobayashi et al.  use new nucleosynthesis yields calculated, from
Z = 0  to $\rm Z=Z_{\odot}$, for supernovae and hypernovae.  With these
new yields, the general trend of O, Na and Al is rather well
reproduced, but not the behaviour of potassium.  The underproduction of
potassium in the models of supernovae has alreay been pointed out by
Samland (\cite{Sam98}).  To solve this problem as well as the
underproduction of Si and Sc, Umeda \& Nomoto (\cite{UN05}) have
proposed to introduce a model of supernova where the density is
assumed to be reduced (during the explosive burning) to increase the
freezeout.

The predictions of Fran\c cois et al.  (\cite {FMC04}) are based on
the yields of Woosley \& Weaver (\cite{WW95}). For a better fit of
the observations they had to systematically increase the production of
K by a factor of 8. In the computations of Kobayashi et al.
(\cite{KUN06}) the ratio [O/Mg] remains stable at very low
metalliciticty. Contrary to that, Fan\c cois et al.  (\cite {FMC04})
predict an increase of [O/Mg] when the metallicity decreases for
$\rm[Mg/H]<-2.5$.  The measurement of [O/Mg] in the two most
metal-poor stars in the Fig.  9 is unfortunately too uncertain (one is
clearly an upper limit) to enable to choose one of the two models.

For Na and Al we give the predictions of Goswami \& Prantzos (\cite {GP00}),
and of Cescutti et al. (private communication).
These predictions
are, as in Fran\c cois et al.  (\cite {FMC04}), based on the yields of
Woosley \& Weaver (\cite{WW95}) but the yields of low mass and 
intermediate mass stars have been adjusted.  Like Kobayashi et al.
(\cite{KUN06}), they predict a decrease of [Na/Mg] and [Al/Mg] when
[Mg/H] decreases, but the agreement with the observations is better
when the model of Kobayashi et al.  is adopted.

\section{Conclusion}

The spectra of the EMP stars, (programme First Stars) analysed previously  
(Cayrel et al. \cite{CDS04}) in LTE have been reanalysed here for Mg and K,
taking into account the departures of LTE.

Abundances of Mg and K:

The NLTE correction for magnesium is in accord with the correction
computed by Gehren and collaborators (2004, 2006) in less metal-poor
stars. The Mg abundance in the giants is raised by a factor of 
about 2 and the NLTE abundance of Mg is the same in giants and in dwarfs. 
The O/Mg ratio is nearly solar (the precision of the O 
 abundance is however rather low).

The NLTE abundance of potassium, computed by adjustement of observed
and computed profiles, has been compared to the work of Takeda et al.
(\cite{TKM09}).  The agreement is generally very good.  When
differences do occur, they arise from the fact that Takeda et al.  used the
equivalent widths given in Cayrel et al.  (\cite{CDS04}) while we
fitted the computed profiles directly to the observed spectra.  (There
are sometimes slight differences in the position of the continuum.)
The NLTE correction for the potassium lines is smaller in the more
metal-poor stars, (as already noted by Takeda et al., \cite{TKM09}).
In Cayrel et al.  (\cite{CDS04}), the
potassium abundance had been roughly corrected for NLTE (by 
applying uniformly a correction of --0.35dex) : the potassium abundances in this 
paper have  therefore been underestimated (especially for those stars that 
are strongly metal-poor).

 Scatters:
 
The scatters of [Mg/Fe] and [K/Fe] are found smaller in the present work,
where departures from LTE are taken into account. The NLTE abundance 
trends are therefore better defined than the previous LTE trends.

The [Mg/H] should be a better reference element than Fe, but even 
whith the new NLTE 
determinations, the scatter of [O/Mg], [Na/Mg]
and [K/Mg] remains (slightly) larger than the scatter of [O/Fe],
[Na/Fe] and [K/Fe]. This fact is surprising, iron being formed in 
processes quite different from those (mainly hydrostatic) supposed to 
be forming O, Mg, Na and K.  An exception is Al : the scatter is the same 
for [Al/Mg] and [Al/Fe].

Trends:

The shapes of the new trends (versus both Fe or Mg) are slightly
different from the LTE trends found previously (Cayrel et al.,
\cite{CDS04}).  For example, the slope of the [K/Fe] ratio versus
[Fe/H], which was slightly positive, becomes slightly negative (Fig.
\ref {abfek}).  Also, the  NLTE values of [Na/Mg], [Al/Mg] and
[K/Mg] show at low metallicity (for $\rm
[Mg/H]<-2.5$) an increase when [Mg/H] decreases.

Comparison with models:

There is some agreement with the models of galactic evolution.
The trends of [O/Mg], [Na/Mg], [Al/Mg] with [Mg/H] in the Galaxy are
rather well represented by the model of Kobayashi et al.
(\cite{KUN06}).  However their model underestimates the production of
potassium.

Finally the precision currently reached in the measurement of high 
resolution high 
signal to noise spectra, obviously deserves careful analyses, using 
accurate atomic parameters,
NLTE computations, and even, in a near future, 3D stellar atmosphere models.

\begin{acknowledgements} SMA kindly acknowledges the support and hospitality 
of the Paris-Meudon
Observatory.  P.B. acknowledges financial support from EU contract
MEXT-CT-2004-014265(CIFIST).  M. S., R. C., F. S., P. B., V. H., P. F.
acknowledge the support of CNRS (PNG and PNPS).
\end{acknowledgements}


\begin{thebibliography}{}

\bibitem[1973]{All73} 
Allen C.W.  1973, Astrophysical Quantities (London: Athlone Press)

\bibitem[2007]{ASK07} 
Andrievsky S.M., Spite M., Korotin S.A., Spite F., Bonifacio P.,
 Cayrel R., Hill V., Fran\c cois P.  2007, A\&A 464, 1081

\bibitem[2008]{ASK08} 
Andrievsky S.M., Spite M., Korotin S.A., Spite F., Bonifacio P.,
 Cayrel R., Hill V., Fran\c cois P.  2008, A\&A 481, 481

 \bibitem[1976]{ALC76} 
Aymar M., Luc-Koenig E., Combet Farnoux F.  1976, JPhB  9, 1279  

\bibitem[2000]{BMB00} 
Ballester P., Modigliani A., Boitiquin O., Cristiani S., Hanuschik R. 
et al.  2000, ESO Messenger 101, 31
 
\bibitem[1973]{BG73} 
Biemont E., Grevesse N.  1973, ADNDT 12, 217

\bibitem[1986]{BB86}
Biemont E., Brault J.W. 1986, Phys. Scr., 34, 751 

\bibitem[2007]{BMS07} 
Bonifacio, P., Molaro, P., Sivarani, T., et 
al.\ 2007, \aap, 462, 851 ({\bf ``First Stars VII''})

\bibitem[2009]{BSC09} 
Bonifacio, P., Spite M., Cayrel R., Hill V., et al.  2009 accepted 
to \aap  (arXiv 0903.4174).

\bibitem[1992]{Bruls92} 
Bruls J.H., Rutten R.J., Shchukina N.  1992, A\&A 265, 237
 
\bibitem[1986]{Carl86} 
Carlsson M.  1986, Uppsala Obs. Rep. 33
 
\bibitem[2004]{CDS04}
Cayrel R., Depagne E., Spite M., Hill V., Spite F., Fran\c cois P., 
Plez B., Beers T.C., Primas F., Andersen J., Barbuy B., Bonifacio P.,
Molaro P., Nordstr\"om B.  2004, A\&A 416, 1117 ({\bf ``First Stars
V''})

\bibitem[2003]{CL03}
Chieffi A.,  Limongi M.  2003 in the ESO astrophysics symposia: 
From Twilight to Highlight: The physics of Supernovae, ed. W. 
Hillebrandt \& B. Leibundgut, p. 367 

\bibitem[2000]{CNZ00}
Chen Y.Q., Nissen P.E., Zhao G., Zhang H.W., Benoni T.  2000, A\&AS 141, 491

\bibitem[2000]{DDK00}
Dekker H., D'Odorico S., Kaufer A., et al. 2000 in Optical and IR 
Telescopes Instrumentation and Detectors, eds I. Masanori \& A.F. 
Morwood Proc. SPIE 4008, 534

\bibitem[1961]{Draw61} 
Drawin H.-W.  1961, ZPhy 164, 513
 
\bibitem[1968]{Draw68} 
Drawin H.-W.  1968, ZPhy 211, 404

\bibitem[2004]{FMC04}
Fran\c cois P, Matteucci F., Cayrel R., Spite M., Spite F., 
Chiappini C.  2004, A\&A 421, 613

\bibitem[2004]{GLS04}
Gehren T., Liang Y.C., Shi J.R.et al.\ 2004, A\&A 413, 1045

\bibitem[2006]{GSZ06}
Gehren T., Shi J.R., Zhang H.W. et al.\ 2006, A\&A 451, 1065

\bibitem[2000]{GP00}
Goswami A., Prantzos N.  2000, A\&A 359, 191

\bibitem[1987]{GS87} 
Gratton R.G., Sneden C.  1987, A\&A 178, 179


\bibitem[2000]{GS00} 
Grevesse N., Sauval A. J., 2000, in Origin of Elements in the Solar
System, Implications of Post-1957 Observations, Proceedings of the
International Symposium.,  Edited by O. Manuel,  Boston/Dordrecht:
Kluwer Academic/Plenum Publishers, p.261

\bibitem[2008]{GEE08}
Gustafsson B., Edvardsson B., Eriksson K., J\o rgensen U. G., Nordlund A., Plez B.
2008, A\&A 486, 951

\bibitem[2002]{HW02}
Heger A., Woosley S.E.  2002, ApJ 567, 532

\bibitem[2008]{HW08}
Heger A., Woosley S.E.  2008, arXiv 0803.3161 (submitted to ApJ)

\bibitem[1979]{Hof79}
Hofsaess D.  1979, ADNDT 24, 285

\bibitem[2000]{IS00}
Ivanova D.V., Shymansky V.V.  2000, ARep 44, 376

\bibitem[1998]{KTN98}
Kobayashi C., Tsujimoto T., Nomoto K., Hachisu I., and Kato M.  1998, ApJ 
503,L155

\bibitem[2006]{KUN06}
Kobayashi C., Umeda H., Nomoto K., Tominaga N., and Ohkubo T.  2006, ApJ 
653,1145

\bibitem[1999]{Kor99} 
Korotin S.A., Andrievsky S.M., Luck R.E.  1999, A\&A 351, 168

\bibitem[1992]{Kur92} 
Kurucz R.L.  1992, The Stellar Population of Galaxies, ed. B. Barbuy, 
A. Renzini, IAU Symp. 149, 225

\bibitem[1993]{Kur93}
Kurucz, R. 1993, ATLAS9 Stellar atmospher Programsand 2km/s
 grid CD-ROM No. 13   Cambridge, Mass.: SAO, 1993, 13
 
\bibitem[1996]{Kur96} 
Kurucz R.L.  1996, Model Atmospheres and Spectrum Synthesis, ed.
S.J. Adelman, F. Kupka, W.W. Weiss, San Francisco, ASP Conf.
Ser. 108, 2
 
\bibitem[1984]{Kur84} 
Kurucz R.L., Furenlid I., Brault J., Testerman L., 1984, Solar Flux
Atlas from 296 to 1300 nm, New Mexico, National Solar
Observatory
 
\bibitem[2003]{Lod03}
Lodders, K. 2003, ApJ, 591, 1220

\bibitem[1980]{MZ80}
Martin W.C., Zalubas, R. 1980, J. Phys. Chem. Ref. Data, 9, 1

\bibitem[2008]{MZG08}
Mashonkina L., Zhao G., Gehren T., Aoki W., Bergemann M., Noguchi K.,
Shi J. R., Takada-Hidai M., Zhang H. W. 2008, A\&A 478, 529

\bibitem[2008]{Mat08}
Matteucci F.  2008, ArXiv 0804.1492v1
 
\bibitem[1969]{MW69}
Miles B.M., Wise W.L.  1969, ADNDT 1, 1

\bibitem[2004]{mskk04}
Mishenina T.V., Soubiran C., Kovtyukh V.V., Korotin S.A.  2004, A\&A 418, 551

\bibitem[1991]{Mor91}
Morton D.C. 1991 Ap.J.Suppl. 77, 119
 
\bibitem[1997]{NHT97}
Nomoto K., Hashimoto M., Tsujimoto T., Thielemann F.-K., Kishimoto 
N., Kubo Y., Nakasato N.  1997, Nucl. Phys. A, 616, 79
 

\bibitem[1989]{PUD89}
Peart B., Underwood J.R.A., Dolder K.  1989, JPhB 22, 1679
 
\bibitem[1995]{PKR95}
Piskunov N.E., Kupka F., Ryabchikova T.A., Weiss W.W., Jeffery C.S.  1995,
A\&AS 112, 525-535
 
\bibitem[1986]{RJL86}
Rahman-Attia, M., Jaouen, M., Laplanche, G., Rachman, A. 1986 J.Phys.B 19,
897
 
\bibitem[1978]{Rut78} 
 Rutten R.J.  1978, SoPh 56, 237

\bibitem[1998]{Sam98}
Samland M.  1998, ApJ 496, 155

\bibitem[1998]{SS98}
Shigeyama T., Tsujimoto T.  1998, ApJ 507, L135

\bibitem[2000]{SMS00}
Shimanskaya N. N., Mashonkina L. I., Sakhibullin N. A.  2000, 
ARep, 44, 530 

\bibitem[1998]{Sch98}
Schoening T., Butler K.  1998, A\&AS 128, 581

\bibitem[1962]{Sea62}
Seaton M.J.  1962, Proc. Phys. Soc. 79, 1105

\bibitem[1981]{SVY81}
Sobelman I.I., Vainshtein L.A., Yukov E.  1981, Excitation of Atoms and
Broadening of Spectral Lines, Springer Ser. in Chem. Phys., Berlin,
Springer

\bibitem[2005]{SCP05}
Spite M., Cayrel R., Plez B., et al.\ 2005, A\&A 430, 655  
({\bf ``First Stars VI''}) 

\bibitem[2006a]{SCH06}
Spite M., Cayrel R., Hill V., et al.\ 2006, A\&A 455, 291
({\bf ``First Stars IX''}) 

\bibitem[2006b]{SCS06}
Spite M., Cayrel R., Spite F., et al.\ 2006, in : Chemical abundances and 
mixing in stars in the Milky Way and its satellites, Proc. 
ESO-Arcetri Workshop, eds. S. Randich \& L. Pasquini, p. 200, Springer

\bibitem[1984]{SH84}
Steenbock W., Holweger H.  1984, A\&A 130, 319

\bibitem[1985]{SC85}
Sugar J., Corliss C.H.  1985, J.PhChRD, 14-2

\bibitem[2009]{TKM09}
Takeda Y., Kaneko H., Matsumoto N., Oshino S., Ito H., Shibuya T. 
2009, in press at PASJ (arXiv0902.4504)

\bibitem[2002]{TZC02}
Takeda Y., Zhao G., Chen Y.Q., Qiu H.M., Takada-Hidai M.  2002, PASJ 54, 275

\bibitem[1995]{TWW95}
Timmes F.X., Woosley S.E., Weaver T.A.  1995, ApJS 98, 617

\bibitem[1970]{TA70}
Truran J.W., Arnett W.D.  1970, ApJ 160, 181

\bibitem[2005]{UN05}
Umeda H., Nomoto K.  2005, ApJ 619, 427

\bibitem[1962]{VReg62}
van Regemorter H.  1962, ApJ 136, 906

\bibitem[1968]{War68}
Warner B.  1968, MNRAS 139, 115

\bibitem[]{WM}
Wise W.L., Martin G.A., NSRDS-NBS 68, P.II

\bibitem[1969]{WSM69}
Wiese W.L., Smith M.W., Miles B.M.  1969, NSRDS-NBS 22 !!!

\bibitem[1995]{WW95}
Woosley S. E., Weaver T. A. 1995, ApJS, 101, 181

\bibitem[2006]{ZGB06}
Zhang H.W., Gehren T., Butler K., Shi J.R., Zhao G.  2006, A\&A 457, 645

\bibitem[2005]{ZZ05}
Zhang H.W., Zhao G.  2005, MNRAS 364, 712

\end{thebibliography}
\end{document}